\documentclass[aas2pp4]{aastex}

\shorttitle{Mass function and binaries in open clusters}
\shortauthors{Sharma et al.}
\begin{document}

\title{Mass functions and photometric binaries in nine\\ open clusters}

\author{Saurabh Sharma\altaffilmark{1}, A. K. Pandey\altaffilmark{1}, K. Ogura\altaffilmark{2}, T. Aoki\altaffilmark{3}, Kavita Pandey\altaffilmark{4},\\T. S. Sandhu\altaffilmark{1,5}  and  R. Sagar\altaffilmark{1}}

\altaffiltext{1}{Aryabhatta Research Institute of Observational Sciences, Nainital, India - 263 129}
\altaffiltext{2}{Kokugakuin University, Higashi, Shibuya-ku, Tokyo 150-8440, Japan}
\altaffiltext{3}{Kiso Observatory, School of Science, University of Tokyo, Mitake-mura, Kiso-gun, Nagano 397-0101, Japan}
\altaffiltext{4}{Department of Physics, DSB Campus, Kumaun University, Nainital, India}
\altaffiltext{5}{Department of Physics, Punjabi University, Patiala 147002, India} 
\date{Received ...../ Accepted .....}

\keywords{Galaxy: open clusters and associations-  Technique: photometric-
Stars: luminosity function, mass function-}

\begin{abstract}

Using homogeneous CCD photometric data from the 105-cm Kiso Schmidt telescope covering a  
$50^\prime \times50^\prime$ field, we study the mass functions (MFs) of nine open clusters. 
The ages and Galactocentric distances of the target clusters vary from 16 - 2000 Myr and 9-10.8 kpc, respectively. 
The values of MF slopes vary from -1.1 to -2.1. 
The classical value derived by Salpeter (1955) for the slope of the IMF is $\Gamma = -1.35$.
The MFs in the outer regions of the clusters are found to 
be steeper than in the inner regions, indicating the presence of mass segregation in the clusters. 
The MF slopes (in the outer region as well as the whole cluster) undergo an exponential decay with 
the evolutionary parameter $\tau$ (= age/ relaxation time). It seems that the evaporation of 
low-mass members from outer regions of the clusters is not significant at larger Galactocentric distances. 
It is concluded that the initial mass function (IMF) in the anticentre direction of the Galaxy might have 
been steeper than the IMF in the opposite direction. A comparison of the observed CMDs of the clusters with 
synthetic CMDs gives a photometric binary content of $\sim$40\%.

\end{abstract}

\maketitle

%________________________________________________________________

\section{Introduction}

The distribution of stellar masses that form in a star-formation event
in a given volume of space is called the Initial Mass Function (IMF) and
together with the star formation efficiency, the IMF dictates the evolution and fate
of star clusters. Present estimations of the observed IMF do not 
constrain the nature of the IMF (see e.g. Kroupa 2007). The universality of the IMF of open clusters 
is still an open question because elementary considerations suggest that
the IMF ought to depend on star-forming conditions (Larson 1998). Therefore 
it is important to find systematic variations of the IMF with different star-forming conditions. 
Identifying these variations would allow us to study early cosmological events (Kroupa 2002).

%Identifying systematic variations of star formation would allow us to understand the physics involved in assembling each of the mass ranges, and thus to probe early cosmological events (Kroupa 2002).

Open clusters  possess many favorable characteristics for IMF studies e.g., clusters 
contain an (almost) coeval set of stars at the same distance with the same metallicity; 
hence, difficulties like complex corrections for stellar birth rates, life times, etc. 
associated with determining the mass function (MF) from field stars are automatically avoided. 
The observed MF of a star cluster can in principle be determined from the observed 
luminosity function (LF) using theoretical stellar-evolutionary models. Since the MFs of 
intermediate/old age open clusters may be affected with time due to stellar as well 
as dynamical evolutionary effects, we can only estimate the present day mass function.  
    
In recent years, luminosity and mass functions have been determined for a number of open 
clusters using homogeneous photoelectric or CCD data and reliable cluster
membership criteria (cf. Piskunov 1976; Sagar et al. 1986, 1988; Scalo 1986, 1998; 
Kjeldsen \& Fransden 1991; Phelps \& Janes 1993; Massey 1995;  Durgapal \& Pandey 2001; 
Pandey et al. 2005, 2007 and references therein). Although the average slope of the MF 
does not seem to be very different from the Salpeter (1955) value, Pandey et al. (2001, 2005) 
found that the nature of the MF in open clusters does not remain the same over the entire 
region of the cluster and the slope of the MF steepens as radial distance from the cluster center increases. 

The nucleus and the corona (extended region of the star cluster) are two main regions in open 
clusters (Kholopov 1969). The nucleus of a cluster usually contains relatively bright i.e. massive 
($\ge$ 3 $M_\odot$) stars along with low mass stars (see e.g. Brandl et al. 1999); whereas the corona, 
which contains a large number of faint and low mass ($\le$ 1 $M_\odot$) stars, has important bearing 
on studies related to the mass function. Extensive studies of the coronal regions of clusters 
have not been carried out so far mainly because of non-availability of photometric data in a large 
field around open star clusters. Therefore, considering the importance of low mass stars 
in the coronae of star clusters, we have generated a homogeneous wide-field photometric data-base 
around 9 open star clusters using a $2K\times2K$ CCD mounted on a Schmidt telescope (Kiso, Japan), 
covering a $\sim 50'\times 50'$ field (Sharma et al. 2006, hereafter Paper I). In this paper 
we study the mass function and photometric binary contents in those nine open clusters. 

\section {Observational data}

Wide field broad-band CCD photometric observations of the clusters were carried out 
during 2001 November 19-25 using the 105 cm Schmidt telescope of the Kiso Observatory. 
A detailed description of the observations and data reduction is given in our previous paper (Paper I). 
Briefly, the CCD camera used a SITe 2048$\times$2048 pixel$^2$ TK2048E chip having a pixel 
size of $24\times24$ $\mu$m. At the Schmidt focus (f/3.1) each pixel of the CCD corresponds 
to $1.^{\prime\prime}5$ and the entire chip covers a field of $\sim 50'\times 50'$ on sky.

Initial processing of the data frames was done in the usual manner using IRAF\footnote{IRAF is distributed by National Optical Astronomy Observatories, USA} and ESO-MIDAS\footnote{ ESO-MIDAS is developed and maintained by the  European Southern Observatory.} data reduction packages. Photometry  of cleaned frames was carried out using 
DAOPHOT Software (Stetson 1987).  The PSF was obtained for each frame using several uncontaminated
stars. The FWHM of the star images varied between $3^{\prime\prime}$ and $4^{\prime \prime}$ from night to night.
The observations were calibrated by observing standard stars in SA95 (Landolt 1992) having 
brightnesses $12.2<V<15.6$ and color indices $0.45<(B-V)<1.51$. Calibration of the instrumental magnitudes 
to the standard system was done using the procedure outlined by Stetson (1992). 
The errors become significantly larger ($\ge$0.1 mag) for stars fainter than $V=20$ mag; therefore, 
the measurements below this magnitude are not reliable and have not been used in the present study.

\section{Luminosity/ mass function}

With the help of color-magnitude diagrams (CMDs) we can derive the observed luminosity function (LF) 
of probable main sequence cluster members and then the mass function (MF) using theoretical evolutionary models, 
for which we adopt those of Bertelli et al. (1994). The factors which influence the determination of LFs 
from the observations are the accuracy of cluster parameters, data incompleteness and field-star 
contamination. The estimation of these factors and their treatment are described in the following subsections.

\subsection{Reddening, distance and age of the clusters}

The cluster parameters, listed in Table 1, are derived using the CMDs as discussed in paper I. 
The CMDs for stars lying within the cluster regions show a well-defined and broad main-sequence (MS). 
Barring Be 62, other clusters manifest a uniform reddening in the cluster region. Since the error in 
magnitude estimation for stars with $V\le$18 mag is $\lesssim 0.05$ mag, we can conclude that 
the presence of probable binaries and field stars should be the main cause for broadening of the 
MS in these clusters. In the case of Be 62 variable reddening in the cluster region along with 
the presence of probable binaries and field stars, should be the cause of the broad MS. 
 
The extinction towards the clusters Be 62, NGC 1960, NGC 2301 and NGC 2323 was estimated 
using the $(U-B)/(B-V)$ two color diagram, whereas in the case of the other five clusters 
NGC 1528, NGC 2287, NGC 2420, NGC 2437 and NGC 2548, the extinction was estimated using 
the $V/(B-V)$ or $V/(V-I)$ CMDs. The reddening in Be 62 varies from $E(B-V)_{min}=0.70$ mag 
to $E(B-V)_{max}=1.00$ mag. The distances and ages of the clusters were obtained by visual 
fitting of the theoretical isochrones by Bertelli et al. (1994) for $Z=0.02$ to the blue 
envelope of the observed MS except in the case of NGC 2420, where we used isochrones for $Z=0.008$,
as Lee et al. (2002) have reported $Z=0.009$ for this cluster. The accuracy of the distance estimates 
is $\sim$ 10 per cent, while that of age determination is about 20 percent. The estimated values 
of $E(B-V)$, distance and age of the target clusters (cf. Paper I) are given in Table 1 and 
have been used in further analysis.  

\subsection{Radial extent of the clusters and field region}

In paper I, we have studied the radial extent and structure of these clusters. The center of the cluster was estimated by convolving a Gaussian kernel with the stellar distribution and taking the point of maximum density as the center. Projected radial stellar density in various concentric circles was obtained by dividing the number of stars in each annulus 
by its area. The extent of the cluster `$r_{cl}$' is defined as the point where the radial density becomes 
constant and merges with the field-star density. Within the uncertainties, the King model (King 1962) reproduces 
well the radial density profiles (RDPs) of the clusters studied in the present work. As an example the RDP of 
the cluster NGC 1960 along with the fitted King profile is shown in Figure 1. For this cluster, the core radius $`r_c$', 
defined as the radial distance at which the value of radial density becomes half of the central 
density (cf. paper I), and cluster extent $`r_{cl}$' come out to be $3.2\pm 0.4$ arcmin (1.2 pc) and 
14 arcmin (5.4 pc), respectively. The structural parameters obtained by fitting the King-model surface-density 
profile to the observed radial density-profile of main-sequence stars having {$V<18$ mag} are taken from paper I and given in Table 2. 

It is well established that clusters have extended regions (coronae). Field-star contamination 
increases considerably in the coronal region of the cluster. The present observations have been made 
in a wide field ($50^\prime \times 50^\prime$); the region outside the cluster extent ($1.5 \times r_{cl}$) 
has been used to estimate the field-star contamination in the cluster region.

\subsection{Probable members and data incompleteness}

To study the LF/MF, it is first necessary to remove field-star contamination from the 
sample of stars in the cluster region. In the absence of a proper-motion study,
we used a statistical criterion to estimate the number of member 
stars in the cluster region. On the basis of a single passband alone it is difficult to establish that a particular star is in fact a member of the cluster. Therefore, two passbands, such as $V$ and $I$, 
are required to identify the cluster members. We used $V/(V-I)$ CMD to estimate the membership as well as 
the LF of the cluster. The contamination due to field stars is greatly reduced by selecting a sample of stars which are located near the well-defined MS as described by Pandey et al. (2001, 2005). The same envelopes were used for the $V/(V-I)$ CMD of the field region to estimate the contamination in the cluster region due to field stars. After normalizing the area we can find the number of field stars which are considered to be present per unit area in each magnitude bin. As an example, selection of the MS sample in the case of NGC 1960 is shown in Figure 2.

The photometric data may be incomplete due to various reasons e.g. crowding of the stars, detection limit etc. 
The incompleteness correction is necessary if we want to analyze the LF/MF of the stars in the cluster. To determine the completeness factor we used the ADDSTAR routine of DAOPHOT II. This method has been used by various authors (cf. Pandey et al., 2005 and references therein). Briefly,  the method consists of randomly adding artificial stars of known magnitude and position into the original frame. The frames are re-reduced using the same procedure used for the original frame. The ratio of the number of stars recovered to those added in each magnitude interval gives the completeness factor, CF, as a function of magnitude.

In practice we followed the procedure given by Sagar and Richtler (1991) and added artificial stars to both $V$ and
$I$ images in such a way that they have similar geometrical locations but differ 
in $I$ brightness according to mean $(V-I)$ colors of the MS stars. The luminosity distribution of 
artificial stars is chosen in such a way that more stars are inserted into the fainter 
magnitudes bins. In all about 15 per cent of the total stars are added so that the crowding 
characteristics of the original frame do not change significantly (cf. Sagar \& Richtler, 1991). 
To have satisfactory statistics for the determination of CF, a number of independent sets 
of artificial stars are inserted into a given data frame (see for e.g. Table 3). The minimum value of the CF of the pair thus obtained is used to correct the data for incompleteness (cf. Sagar \& Richtler, 1991). 
As an example, the CF along with relevant information for NGC 1960 is given in Table 3. As expected the incompleteness of the data increases with increasing magnitude and increasing stellar crowding.

The number of probable cluster members in the two sub-regions of the cluster 
was obtained by subtracting the contribution of field stars (corrected for data
incompleteness) from each magnitude bin of the contaminated sample of MS stars
(also corrected for data incompleteness). The statistics in the case of one of the clusters, NGC 1960, is given in Table 4.

\section{Results}

\subsection{ Mass function}

The MF is often expressed by the power law,
$N (\log m) \propto m^{\Gamma}$, where the slope of the MF is given as:
$$ \Gamma = d \log N (\log m)/d \log m  $$
\noindent
 where $N  (\log m)$ is the number of stars per unit logarithmic mass interval.
The classical value derived by Salpeter (1955) for the slope of the IMF is $\Gamma = -1.35$. 
The main-sequence LF, obtained with the help of CMDs for two sub-regions of the target clusters, has been 
converted into a MF using the theoretical model of Bertelli et al. (1994). The resultant MF data 
for the cluster NGC 1960 are given in Table 5. The MFs of the target clusters for two sub-regions as well as for the whole cluster region are shown in Figure 3. In the specified mass range the MF can be represented by a single power law. The value of MF slopes $\Gamma$, obtained by using the least-square solution in the specified mass range has also been given in Table 6. For intermediate age clusters the mass ranges are in general $\sim 1- 3 M_\odot$ but for Be 62 and NGC 1960, which are the youngest clusters in the sample, the mass range is slightly higher. 
Despite large errors in $\Gamma$ values, Table 6 indicates that in seven out of nine clusters
the values of $\Gamma$ are steeper in the outer region as compared to that in the inner region.
Barring the cluster NGC 1960, the difference in the values of $\Gamma$ for inner and outer regions
is less than $3~\sigma$ ($\sim 1-2~\sigma$). The steeper values of $\Gamma$ in the outer regions 
may be attributed to mass segregation. There is evidence of mass segregation in some Galactic 
and LMC clusters, with higher mass stars preferentially located
towards the center of the cluster (see e.g. Fisher et al. 1998; Pandey et al. 1992, 2001, 2005 and references therein, Kumar et al. 2008). 

To evaluate the degree of mass segregation in clusters, we subdivided the samples into two mass groups as indicated in Figure 4, which shows the cumulative distribution of MS stars as a function of radius in two different mass groups.
In the case of six clusters (Be 62, NGC 1528, NGC 1960, NGC 2323, NGC 2420 and NGC 2437) 
Figure 4 reveals the effect of mass segregation in the sense that relatively massive stars tend to 
lie near the cluster center. 
In the case of Be 62, NGC 1960, NGC 2323 and NGC 2420 the Kolmogrove-Smirnov test confirms the above mentioned
mass segregation at a confidence level better than 99 per cent, whereas in the case of NGC 1528 and NGC 2437
the confidence level is better than 90 per cent and 95 per cent, respectively.

\subsection{Dynamical state of the clusters}

Observations of mass segregation in several young clusters in the Galaxy 
(e.g. Moffat 1970; Herbst \& Miller 1982; Larson 1982; Sagar et al. 1988; 
Pandey, Mahara \& Sagar 1992; Hillenbrand 1997; Raboud \& Mermilliod 1998) 
as well as in the Magellanic Clouds (Fischer et al. 1998 and references therein) suggest that mass 
segregation may be the imprint of the star formation process itself. On the other hand, if clusters
had a uniform spatial stellar-mass distribution at the time of formation, the spatial stellar-mass 
distribution would change with time as clusters evolve dynamically. Because of equipartition of 
energy the low-mass stars would attain high velocity and move away from the cluster center, 
consequently higher concentration of high mass stars towards the center of the cluster could be
observed (cf. Mathieu 1985; Mathieu \& Latham 1986;
McNamara \& Sekiguchi 1986). To decide whether mass segregation is primordial or due to dynamical 
relaxation, we have to estimate the dynamical relaxation time, $T_E$, the time in which the 
individual stars exchange sufficient energy so that their velocity distribution approaches 
that of a Maxwellian equilibrium. The dynamical relaxation time is given by:

\bigskip
{\large
~~~~~~~~~~~~~~~$T_{E} =  \frac{8.9\times10^5 N^{1/2} R_h^{3/2}}{\bar{m}^{1/2} log (0.4N)} $} $^,$

\bigskip
\noindent
where N is the number of cluster stars, $R_{h}$ is the radius containing half of
the cluster mass and $\bar{m}$ is the average mass of cluster stars (Spitzer \& Hart 1971). 

We have estimated the relaxation time $`T_E$' for all the target clusters to 
decide whether the mass segregation discussed above is primordial or due to dynamical relaxation. The total number of MS stars and the total mass of the MS stars in the given mass range (see Table 7) are obtained with the help of the MF.
This mass should be considered as a lower limit to the total mass of the cluster. 
For the half-mass radius, we used half of the cluster extent ($R_{cl}$) obtained from the
optical data (see. Table 2). The cumulative distribution of all the stars, shown by dotted curve
in Figure 4, indicates that the 50 per cent of the cluster stars lie within $\sim 0.48\pm 0.10~ R_{cl}$,
therefore half of the cluster extent seems to be a reasonable approximation for the 
half-mass radius. The values of various parameters as well as the resultant $T_E$ for the target clusters 
are given in Table 7. 

A comparison of cluster age with its dynamical relaxation time in the case of intermediate/old age 
clusters (age $>10^8$ yr; see Table 7) indicates that the former is greater than the latter, 
leading to the conclusion that dynamical evolution could also be the reason for the observed mass segregation. 
In the case of the young clusters Be 62 and NGC 1960, the dynamical relaxation time is comparable 
to the age of the clusters hence the  observed segregation in these clusters could be because of both the imprint of
the star formation process and dynamical relaxation.

\subsection {Synthetic CMDs} 

During the last decade, synthetic CMDs have been used to study various properties
of clusters, e.g., the MF and the influence of unresolved photometric
binaries on the LF, etc. (cf. Sandhu et al. 2003 and references therein). By comparing the synthetic integrated 
luminosity-function (ILF) and synthetic color-distribution with the corresponding observed distribution,
these authors estimated the photometric binary content in three intermediate-age open clusters. 
Following the procedure of Sandhu et al. (2003) we calculated the ILF and 
$\Delta(V-I) = (V-I)_* - (V-I)_{MS}$ for each star, where  $(V-I)_*$ is the observed color of a star 
and $(V-I)_{MS}$ is the corresponding color of the MS. The $\Delta(V-I)$ frequency distribution of stars, in the specified magnitude range (cf. Table 9) of the statistically cleaned CMD was compared with 
the $\Delta(V-I)$ frequency distribution of the synthetic CMDs (for details see Sandhu et al. 2003). 
Statistically cleaned CMDs were obtained using the following statistical procedure. For a star in the $V, (V-I)$ CMD of the field region, the nearest star in the cluster's $V,(V-I)$ CMD within $V\pm0.25$ and $(V-I)\pm0.13$ of the field star was removed. While removing the stars from the cluster CMD, the number of stars in each luminosity bin was maintained as per the completeness corrected LF (cf. Sect. 3.3). 

Figure 5 shows the statistically cleaned CMD for the cluster NGC 1960. The CMDs of the clusters in 
the anticentre direction of the Galaxy are strongly affected by the background population of the 
Norma-Cygnus arm (see e.g. Pandey et al. 2006). The effect of the background population can even
be seen in the statistically-cleaned CMDs. 
Therefore to avoid contamination due to background population we limit the MS population towards the fainter
end and the fainter end limit is given in Table 9.

Figure 6 shows the comparison of the observed ILF of the cluster NGC 1960 with the best-fit synthetic 
ILF for various percentages of photometric binary content along with the obtained value of the MF slope. 
We assumed that the mass-ratio (mass of secondary / mass of primary) varies in the range of $0.75-1.0$.
The results for the target clusters are given in Table 8, which indicate that the value of the MF slope, 
$\Gamma$, for the observed MF is in agreement (within the errors) with the value of the MF slope obtained 
for the synthetic CMDs without binary content. Table 8 also indicates that the {\it true/intrinsic} value of 
`$\Gamma$' becomes steeper if the photometric binary fraction is higher. This result is in  agreement with that obtained by Sandhu et al. (2003) in the case of three open clusters.

The comparison of observed and synthetic distributions along with $\chi^2$
values is shown as an example in Figures 7 and 8 for the clusters NGC 1960 and NGC 1528, respectively. 
The results are given in Table 9 which indicate an average detectable photometric 
binary content of $\sim$ 30\% - 40\% in the present sample. 
Mermilliod \& Mayor (1989) found $25 \% - 33 \%$ spectroscopic binaries in open clusters.
Aparicio et al. (1990) and Durgapal et al. (2001) reported $>25\%$ and $10\%-20\%$ photometric 
unresolved binaries in the clusters. In case of the Pleiades cluster, Bouvier et al. (1997)
reported a binary (wider,visual) frequency of about $28\pm4\%$ for G and K dwarfs. Using the
infrared speckle observations of the Hyades cluster, Patience et al. (1998) found that $\sim40\%$
stars are binary. Mason et al. (1998) have estimated a binary fraction
(both spectroscopic/unresolved as well as visual) of $75\%$ in clusters/associations,
whereas Jeffries et al. (2001), for the cluster NGC 2516, found a photometric binary fraction of $26 \pm 5 \%$
for A to M-type systems with mass ratio between 0.6 and 1.

\section{Discussions}

The mass functions of two clusters, NGC 1960 and NGC 2323 are significantly steeper ($>3~\sigma$ level)
than the Salpeter value, whereas the MFs of the clusters NGC 1528, Be 62 and NGC 2437 are 
found to be steeper but with lower significance level (2.1, 1.6 and 1.6 $\sigma$ level).
The MF slopes for the youngest clusters Be 62 and NGC 1960 are based on a wide range of mass, 
i.e. 1.1-11.2 $M_\odot$ and 1.0-6.8 $M_\odot$, respectively, while for other clusters the MF is 
derived for a relatively narrow mass range. The mass function slopes for outer regions are 
always steeper than the slopes for the inner regions. 
Mass function slopes of the youngest clusters, namely Be 62 and NGC 1960, are found to be $-1.88\pm0.34$ 
and $-1.80\pm0.14$, respectively, which are comparable to the slopes obtained for intermediate/old clusters 
of the sample. The MF slopes for NGC 2287, NGC 2301, NGC 2420 and NGC 2548 are found to be 
comparable to the Salpeter value (-1.35). Bonatto \& Bica (2005) have estimated  $\Gamma$ values of 
$-1.5\pm 0.2$ and $-1.3 \pm 0.2$ for the clusters NGC 2287 and NGC 2548, respectively. 
For cluster NGC 1528, in the mass range 1.12-2.82 $M_\odot$, Francic (1989) has estimated a steeper 
mass-function slope ($-2.78\pm0.31$). The present study also indicates a steeper mass function 
slope for NGC 1528. In the case of NGC 2323, for the mass range 0.40 - 3.90 $M_\odot$, 
Kalirai et al. (2003) have reported $\Gamma =-1.94\pm0.15$ which agrees well with the present 
value ($-2.01 \pm 0.17$). For NGC 1960, the present mass-function slope is steeper than the value 
given by Sanner et al. (2000) i.e. $-1.23\pm0.17$ in the mass range 0.72-9.4 $M_\odot$.

In order to investigate the relationship between relaxation time and cluster age with dynamical evolution, and to estimate the corresponding effects on MFs we calculate for each cluster the evolutionary parameter, $\tau$, which is defined as the ratio of the cluster age to the relaxation time, $\tau = age/T_E$. Table 7 lists the estimated values of $\tau$  for each cluster.  Figure 9 shows $\Gamma$ as a function of $\tau$. The slope of the MFs for six clusters, obtained from synthetic CMDs for 0\% binary are shown by open circles. Data for two clusters (NGC 1907 and NGC 1908) have been taken from Pandey et al. (2007) and shown by triangles. 
Although the errors in $\Gamma$ values are large, Figure 9 clearly shows a systematic decreasing trend in $\Gamma$
with $\tau$, particularly in the outer regions of the clusters, indicating a exponential decay of $\Gamma$ with $\tau$.
Bonatto \& Bica (2005) and Maciejewski \& Niedzieski (2007) have also concluded the same. However, the dependence  of $\Gamma$ in the central region of the cluster on $\tau$ does not show the same trend.
The decreasing trend may be represented by an exponential of the form:

$ \Gamma =  \Gamma_0 + e^{a/\tau} $ 

\noindent with $\Gamma_0 = -1.9 \pm 0.10$, $a = -17.6 \pm 7.4$ (correlation coefficient $\sim 0.8$, 
reduced $\chi ^2=0.04$) and $\Gamma_0 = -2.11 \pm 0.10$, $a = -18.6 \pm 7.9$ (correlation coefficient 
$\sim 0.8$, reduced $\chi^2=0.04$) 
for the whole cluster and outer regions, respectively, which indicate that the deceasing trend 
of $ \Gamma$ with $\tau$ in the whole/outer cluster region is significant.  
Whereas for the inner region we estimate $\Gamma_0 = -2.27 \pm 0.18$, $a = 0.11 \pm 0.25$ 
(correlation coefficient $\sim 0.5$, reduced $\chi^2=0.22$) which indicates no correlation between
$\Gamma$ and $\tau$ in the inner region.

The parameter $\Delta \Gamma = \Gamma_{inner} -\Gamma_{outer}$, where $\Gamma_{inner}$ and $\Gamma_{outer}$ are the MF slopes for the inner and outer regions, respectively, can reveal information about mass segregation. 
Maciejewski \& Niedzieski (2007) reported no correlation between $\Delta \Gamma$ and $\tau$ but $\Delta \Gamma$ 
increases with age in the case of clusters older than $\sim$ 100 Myr. In Figure 10 we plot $\Delta \Gamma$ as a 
function of $\tau$ and age, which does not reveal any relation between $\Delta \Gamma$ and age of the cluster, however it seems that $\Delta \Gamma$ decreases systematically with increase in $\tau$. 
The decrease in $\Delta \Gamma$ with $\tau$ can be interpreted as evaporation of low mass stars from the outer region.
Figure 10 also indicates a systematic variation of $\Delta \Gamma$  
as a function of Galactocentric distance, in the sense that $\Delta \Gamma$ increases with increase in the 
Galactocentric distance, indicating that evaporation of low-mass members 
from outer region of the clusters is not significant at larger Galactocentric distances. 
Here we would like to point out that a larger sample is needed to get a conclusive
view about variation of $\Delta \Gamma$ with the $\tau$, age and Galactocentric distance.

To study the dependence of the MF on the core radius $r_c$, cluster extent $r_{cl}$, Galactocentric 
distance $R_G$ and age of the star cluster, we used the values derived in paper I. 
To convert the distance to Galactocentric distance, the Galactocentric distance of the Sun 
is taken as 8.5 kpc (Allen 2000). Figure 11 shows the dependence of cluster MF slope 
$\Gamma$ on $r_c$, $r_{cl}$, $R_G$ and cluster age. The Salpeter value for $\Gamma$ (-1.35) 
is shown as a straight line. 
Figure 11 indicates that clusters having core radii greater than $\sim1$ pc and cluster radii greater than
$\sim4$ pc have steeper MF than the Salpeter MF. The difference varies from $1.6-2.1~\sigma$ (Be 62,
NGC 2437 and NGC 1528) to $>3~\sigma$ (NGC 1960 and NGC 2323). The cluster located
at $R_G\sim9.5-10$ kpc also show a steeper MF at $1.6~\sigma$ to $3.2~\sigma$ level.
As these clusters are situated in the anticentre direction of the Galaxy, it can be suggested that the 
IMF might have been steeper towards the anti-center direction as compared to other directions in the Galaxy. 
$\Gamma$ does not show any trend with the ages of the clusters.

\section{Summary and conclusion} 

In this paper we studied MFs of nine open star clusters located in the anti-center direction of 
the Galaxy, using wide field CCD photometric data taken from the Kiso Schmidt telescope. The values of MF slopes vary from -1.1 to -2.1. The main conclusions of the study are as follows.

\begin{itemize}

\item [(1)] The youngest clusters Be 62 and NGC 1960 in the present sample have steeper MF slopes, $-1.88 \pm 0.34$ and $-1.80 \pm 0.14$ respectively, than the Salpeter value. The observed MF of Be 62 can be assumed  as the initial mass function (IMF), since the dynamical relaxation time is longer than the age of Be 62. 

\item [(2)] Three intermediate-age clusters (NGC 1528, NGC 2323 and NGC 2437) have steeper mass-function 
slopes, whereas four other intermediate-age  clusters have MF slopes comparable to the Salpeter value.

\item[(3)] Most of the clusters of the present sample show the effect of mass segregation. Mass segregation 
in the case of the young cluster Be 62 indicates that mass segregation could be due to the star formation process itself, whereas in the case of intermediate/old  clusters the mass segregation can also be explained on the 
basis of dynamical evolution.

\item [(4)] The MF slope of the outer region/whole cluster region is seem to be related to the dynamical-evolution parameter $\tau$. The MF slopes (particularly in the outer region of the cluster) undergo an exponential decay with $\tau$.

\item [(5)] There is evidence for initial mass-segregation within the young clusters and decrease in $\Delta \Gamma$ with $\tau$ is interpreted as evaporation of low mass stars from the outer regions of the clusters.

\item [(6)] It is found that evaporation of low-mass members from outer regions of the clusters is not significant at larger Galactocentric distances. 

\item [(7)] The clusters having larger core/cluster radii have relatively steeper MF slopes. At larger Galactocentric distances the MFs of the clusters are found to be steeper. We do not find any correlation between MF and age of the clusters.

\item [(8)] The present analysis  of the synthetic CMDs reveals a detectable photometric binary content of about 30\% - 40\% in the intermediate age clusters.

\end{itemize}

\section{acknowledgments}

Authors are thankful to referee Prof. A. Moffat for useful comments which
improved the contents of the paper.
AKP is thankful to DST (India) and JSPS (Japan) for providing funds to
visit KISO Observatory to carry out the observations and to the staff of KISO
Observatory for their generous help during his stay.

\clearpage
\newpage

\begin{table*}
\centering
\begin{minipage}{140mm}
\caption{Parameters of the target clusters taken from paper I.
To determine the Galactocentric distances $R_G$ to the clusters, a value of 8.5 kpc (Allen 2000) has 
been assumed for the Galactocentric distance of the Sun.\label{Table: 8}}
\begin{tabular}{@{}r r r c c c r@{}}
\hline
Cluster &$\alpha_{2000}$ &$\delta_{2000}$ &$E(B-V)$ & Log age &Distance &$R_G$ \\
        &(h:m:s)&(d:m:s)& (mag)    &  (yr)      & (kpc) & (kpc)\\

\hline
Be 62   &$ 01:01:15.5$&$  63:56:17  $&0.70-1.00&7.2 &2.32&9.98 \\
NGC 1528&$ 04:15:24.2$&$  51:15:23  $&0.26&8.6 & 1.09&9.48 \\
NGC 1960&$ 05:36:20.8$&$  34:08:31  $&0.22&7.4 & 1.33&9.82 \\
NGC 2287&$ 06:45:58.7$&$ -20:44:09  $&0.01&8.4 & 0.71&8.96 \\
NGC 2301&$ 06:51:46.4$&$  00:27:30  $&0.03&8.2 & 0.87&9.25 \\
NGC 2323&$ 07:02:47.4$&$ -08:20:43  $&0.20&8.0 & 0.95&9.23 \\
NGC 2420&$ 07:38:24.8$&$  21:34:30  $&0.04&9.3 & 2.48&10.76\\
NGC 2437&$ 07:41:58.1$&$ -14:49:28  $&0.10&8.4 & 1.51&9.51 \\
NGC 2548&$ 08:13:42.9$&$ -05:46:37  $&0.03&8.6 & 0.77&9.02 \\
\hline
\end{tabular}
\end{minipage}
\end{table*}

\begin{table*}
\centering
\begin{minipage}{140mm}
\caption{Structural parameters of the target open clusters taken from paper I.\label{Table: 2}}
\begin{tabular}{@{}r r r r r r r@{}}
\hline
&&& \multicolumn{2}{c}{Optical} &\multicolumn{2}{c}{2MASS} \\
Cluster &l &b &Core radius &Cluster extent &Core radius& Cluster extent \\
 &(degree)&(degree)&arcmin (pc)&arcmin (pc)&arcmin (pc)&arcmin (pc)\\
\hline

Be 62   & $ 123.98$&$  1.10$&2.2 (1.5)&10 (7)&2.5 (1.7)&12 (8)\\
NGC 1528& $ 152.06$&$  0.26$&8.3 (2.6)&15 (5)&18.5 (5.9)&24 (8)\\
NGC 1960& $ 174.52$&$  1.07$&3.2 (1.2)&14 (5)&3.8 (1.5)&21 (8)\\
NGC 2287& $ 231.02$&$-10.44$&1.4 (0.3)&12 (3)&12.7 (2.6)&16 (3)\\
NGC 2301& $ 212.56$&$  0.28$&1.9 (0.5)&9 (2) &4.5 (1.1)&20 (5)\\
NGC 2323& $ 221.67$&$ -1.33$&6.5 (1.8)&17 (5)&6.7 (1.9)&22 (6)\\
NGC 2420& $ 198.11$&$ 19.63$&1.4 (1.0)&10 (7)&1.3 (0.9)&9 (7)\\
NGC 2437& $ 231.86$&$  4.06$&6.8 (3.0)&20 (9)&9.6 (4.2)&25 (11)\\
NGC 2548& $ 227.87$&$ 15.39$&1.5 (0.3)&8 (2) &2.4 (0.5)&8 (2)\\
\hline
\end{tabular}
\end{minipage}
\end{table*}

\begin{table*}
\centering
\begin{minipage}{140mm}
\caption{Variation of the completeness factor with MS brightness in different radial ($r, ~in~ arcmin$)
regions for the cluster NGC 1960. $N_f$ is the number of frames generated; $nc_a$,$ ncn_a$,$nf_a$, $nc_r$,$ncn_r$,$nf_r$ 
and $cft_c$,$cft_{cn}$, $cft_f$ are numbers of star added, recovered and corresponding completeness 
factors in inner, outer and field regions respectively.}
\begin{tabular}{@{}rr|rrr|rrr|rrr@{}}
\hline
\hline
Range in mag &$N_f$&\multicolumn{3}{c}{ Inner region} & \multicolumn{3}{c}{Outer region} &\multicolumn{3}{c}{ Field region} \\
  &&\multicolumn{3}{c}{ ($r<3^\prime.2$)} & \multicolumn{3}{c}{($3^\prime.2 \leq r<14^\prime$)} &\multicolumn{3}{c}{ ($17^\prime.5<r<22^\prime.5$)} \\

          && $nc_a$ & $nc_r$&$cft_c$&$ ncn_a$&$ncn_r$&$cft_{cn}$& $nf_a$& $nf_r$& $cft_f$\\
\hline

$V$ band:\\
13.6-14.6 & 4&    19&    19&  1.00&   240&   237&  0.99&    277&   274&  0.99   \\
14.6-15.6 & 7&    30&    30&  1.00&   480&   467&  0.97&    552&   542&  0.98   \\
15.6-16.6 &10&    53&    46&  0.87&   740&   723&  0.98&    892&   878&  0.98   \\
16.6-17.6 &13&    62&    58&  0.94&  1153&  1109&  0.96&   1268&  1243&  0.98   \\
17.6-18.6 &17&    83&    73&  0.88&  1674&  1549&  0.93&   1841&  1739&  0.94   \\
18.6-19.6 &21&   153&   118&  0.77&  2462&  2229&  0.91&   2662&  2475&  0.93   \\
                                                                  
$I$ band:\\
13.6-14.6 & 4&    16&    16&  1.00&   256&   255&  1.00&    293&   291&  0.99   \\
14.6-15.6 & 7&    27&    26&  0.96&   528&   520&  0.98&    599&   592&  0.99   \\
15.6-16.6 &10&    54&    49&  0.91&   808&   789&  0.98&    920&   909&  0.99   \\
16.6-17.6 &13&    71&    63&  0.89&  1168&  1122&  0.96&   1368&  1332&  0.97   \\
17.6-18.6 &17&    88&    76&  0.86&  1743&  1626&  0.93&   1888&  1778&  0.94   \\
18.6-19.6 &21&   143&   110&  0.77&  2609&  2310&  0.89&   2802&  2565&  0.92   \\

\hline
\end{tabular}
\end{minipage}
\end{table*}

\begin{table*}
\centering
\begin{minipage}{140mm}
\caption{Luminosity function for the two sub-regions and the whole cluster region in the case of NGC 1960.
$n_{c}$, $n_{f}$ are the numbers of stars (corrected for data incompleteness) 
in the sub-regions of the cluster and expected in the field,
respectively, and $n_{p}$ is the number of probable cluster members. }
\begin{tabular}{@{}rrrrrrrr@{}}
\hline
\hline
 Range&\multicolumn{3}{c}{Inner region}&\multicolumn{3}{c}{Outer region}&Whole region \\
V mag&$n_{c}$&$n_{f}$&$n_{p}$& $n_{c}$&$n_{f}$&$n_{p}$& $n_{p}$  \\
\hline

09.6-10.6&   2&     0&     2&     5&     2&     3&       5\\
10.6-11.6&   7&     0&     7&    12&     1&    11&       18\\
11.6-12.6&  10&     1&     9&    24&    11&    13&       22\\
12.6-13.6&  11&     1&    10&    46&    23&    23&       33\\
13.6-14.6&  16&     2&    14&    71&    45&    26&       40\\
14.6-15.6&  19&     5&    14&   133&    89&    44&       58\\
15.6-16.6&  17&    10&     7&   244&   190&    54&       61\\

\hline
\end{tabular}
\end{minipage}
\end{table*}

\begin{table*}
\centering
\begin{minipage}{140mm}
\caption{Mass function of the cluster NGC 1960. The number of probable cluster
members ($N$) was obtained after subtracting the expected contribution of
field stars in each magnitude range. log $\phi$ represents log (d$N$/dlog $m$).}
\begin{tabular}{@{}rrrrrrrrr@{}}
\hline
\hline

Range&Mass&Mean&\multicolumn{2}{c}{Inner region}&\multicolumn{2}{c}{Outer region}&\multicolumn{2}{c}{Whole
region}\\
V (mag)&$(M_{\odot})$&log $M_{\odot}$& N&log $\phi$& N& log $\phi$&N& log $\phi$ \\ \hline

 9.6-10.6& 6.82-4.97&   0.7703&    2&  1.1625&     3&  1.3386&     5&   1.5604\\
10.6-11.6& 4.97-3.45&   0.6241&    7&  1.6458&    11&  1.8421&    18&   2.0559\\
11.6-12.6& 3.45-2.35&   0.4621&    9&  1.7303&    13&  1.8900&    22&   2.0983\\
12.6-13.6& 2.35-1.74&   0.3102&   10&  1.8861&    23&  2.2478&    33&   2.4046\\
13.6-14.6& 1.74-1.42&   0.1985&   14&  2.2004&    26&  2.4692&    40&   2.6563\\
14.6-15.6& 1.42-1.18&   0.1136&   14&  2.2393&    44&  2.7367&    58&   2.8566\\
15.6-16.6& 1.18-1.01&   0.0385&    7&  2.0085&    54&  2.8958&    61&   2.9487\\
\hline
\end{tabular}
\end{minipage}
\end{table*}

\begin{table*}
\centering
\begin{minipage}{140mm}
\caption{Mass-function slope $\Gamma$ for two sub-regions and for the whole cluster region
in the given mass range. $\sigma$ is the standard deviation of the slopes.\label{Table: 8}}
\begin{tabular}{@{}r c c r r @{}}
\hline
Cluster &Mass range& \multicolumn{3}{c}{Mass function slopes $(\Gamma \pm \sigma)$} \\
	&$(M_\odot)$&Inner region & Outer region & Whole cluster\\
\hline
Be 62     &        11.17-1.14 & $ -0.89\pm0.17$ &$ -2.10\pm0.74 $&$ -1.88\pm0.34$\\
NGC 1528  &         2.55-0.73 & $ -1.96\pm0.42$ &$ -2.17\pm0.43 $&$ -2.10\pm0.35$\\
NGC 1960  &         6.82-1.01 & $ -1.25\pm0.24$ &$ -1.99\pm0.15 $&$ -1.80\pm0.14$\\
NGC 2287  &         2.70-0.83 & $ -1.35\pm0.86$ &$ -1.22\pm0.27 $&$ -1.22\pm0.19$\\
NGC 2301  &         2.78-0.82 & $ -0.85\pm0.33$ &$ -1.56\pm0.54 $&$ -1.34\pm0.32$\\
NGC 2323  &         4.22-0.67 & $ -1.69\pm0.09$ &$ -2.28\pm0.31 $&$ -2.01\pm0.17$\\
NGC 2420  &         1.44-0.67 & $ -0.93\pm0.32$ &$ -1.50\pm0.56 $&$ -1.30\pm0.39$\\
NGC 2437  &         3.51-1.02 & $ -1.72\pm0.13$ &$ -2.30\pm0.62 $&$ -2.03\pm0.42$\\
NGC 2548  &         2.46-0.82 & $ -1.11\pm0.85$ &$ -1.02\pm0.36 $&$ -1.12\pm0.70$\\

\hline
\end{tabular}
\end{minipage}
\end{table*}

\begin{table*}
\centering
\begin{minipage}{140mm}
\caption{Various parameters for the target clusters in the given mass range (see Table 6) 
used for calculating the dynamical-evolution time $T_E$.\label{Table: 8}}
\begin{tabular}{@{}r c c c c c c c c @{}}
\hline

Cluster   &Number of &   Half mass  &Total mass &Average mass& Age & Dynamical time & $\tau$\\
          &  stars   &   radius $R_h$ (pc)& ($M_\odot$) &  $\bar{m}$ ($M_\odot$) &(Myr)& $T_E$ (Myr)& ($age/T_E$)\\
\hline  
Be 62     &     143   &     3.4   &  312   &  2.18    &       16   &   26& 0.6\\
NGC 1528  &     465   &     2.4   &  530   &  1.14    &       400  &   29& 13.8\\
NGC 1960  &     232   &     2.7   &  431   &  1.86    &       25   &   22& 1.1 \\
NGC 2287  &     102   &     1.2   &  142   &  1.39    &       250  &   6 & 41.7 \\
NGC 2301  &     120   &     1.1   &  165   &  1.38    &       160  &   6 & 26.6 \\
NGC 2323  &     695   &     2.3   &  803   &  1.16    &       100  &   31& 3.2 \\
NGC 2420  &     752   &     3.6   &  607   &  0.81    &       2000 &   75& 26.7 \\
NGC 2437  &     647   &     4.4   &  1026  &  1.59    &       250  &   69& 3.6 \\
NGC 2548  &      45   &     0.9   &  61    &  1.36    &       400  &   4 & 100.0     \\
\hline
\end{tabular}
\end{minipage}
\end{table*}

\begin{table*}
\centering
\begin{minipage}{140mm}
\caption{Slope of the mass function $\Gamma$ obtained from synthetic CMDs
for various percentages of binary content. \label{Table: 9}}
\begin{tabular}{@{}c c r c r r r @{}}
\hline
Binary \% &  NGC 1528 & NGC 1960 & NGC 2287 & NGC 2301 & NGC 2323 & NGC 2420 \\
\hline
 0  & $-2.02\pm0.13$ & $-1.85\pm0.12$ & $-1.31\pm0.21$ & $-1.32\pm0.16$ & $-2.10\pm0.13$ & $-1.06\pm0.19$ \\
10  & $-2.14\pm0.20$ & $-1.79\pm0.16$ & $-1.36\pm0.18$ & $-1.57\pm0.17$ & $-2.16\pm0.10$ & $-1.70\pm0.27$ \\
20  & $-2.43\pm0.17$ & $-2.01\pm0.11$ & $-1.51\pm0.18$ & $-1.65\pm0.15$ & $-2.27\pm0.17$ & $-1.92\pm0.41$ \\
30  & $-2.51\pm0.23$ & $-2.28\pm0.18$ & $-1.52\pm0.15$ & $-1.72\pm0.11$ & $-2.37\pm0.13$ & $-2.20\pm0.45$ \\
40  & $-2.67\pm0.30$ & $-2.29\pm0.20$ & $-1.62\pm0.28$ & $-1.88\pm0.25$ & $-2.53\pm0.13$ & $-2.48\pm0.68$ \\
50  & $-2.68\pm0.26$ & $-2.29\pm0.20$ & $-1.70\pm0.16$ & $-1.90\pm0.19$ & $-2.66\pm0.22$ & $-2.58\pm0.59$ \\
\hline
\end{tabular}
\end{minipage}
\end{table*}

\begin{table*}
\centering
\begin{minipage}{140mm}
\caption{Binary fraction in various clusters. The expected error in estimation of binary content is $\sim10$ per cent. \label{Table: 10}}
\begin{tabular}{@{}c c c c @{}}
\hline
Cluster  &$V$ range& Mass-range   & Photometric binary\\
         &  (mag)  & ($M_\odot$)  & content (\%) \\
\hline
NGC 1528 & 11-17   &   0.8 -3.0   &       40\\
NGC 1960 & 10-17   &   0.9 -6.1   &       30\\
NGC 2287 & 09-16   &   0.8 -3.2   &       30\\
NGC 2301 & 10-17   &   0.7 -3.1   &       40\\
NGC 2323 & 10-17   &   0.8 -3.9   &       75\\
NGC 2420 & 15-19   &   0.7 -1.3   &       40\\
\hline
\end{tabular}
\end{minipage}
\end{table*}

\clearpage

%-------------Fig 1----------------------------------------------------------------
\begin{figure*}
\centering
\includegraphics[height=5cm,width=8cm]{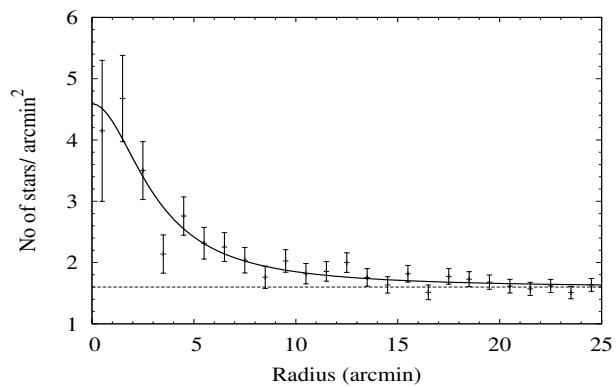}
\caption{ The variation of stellar surface density for stars having $V\le18$, as a function of radial distance in NGC 1960. Continuous curve shows a least square fit of the King (1962) profile to the observed data points. The error bars represents $\pm \sqrt{N}$ errors. The dashed line
indicates the density of field stars.}
\end{figure*}

%-------------Fig 2----------------------------------------------------------------
\begin{figure*}
\centering
\includegraphics[height=7cm,width=8cm]{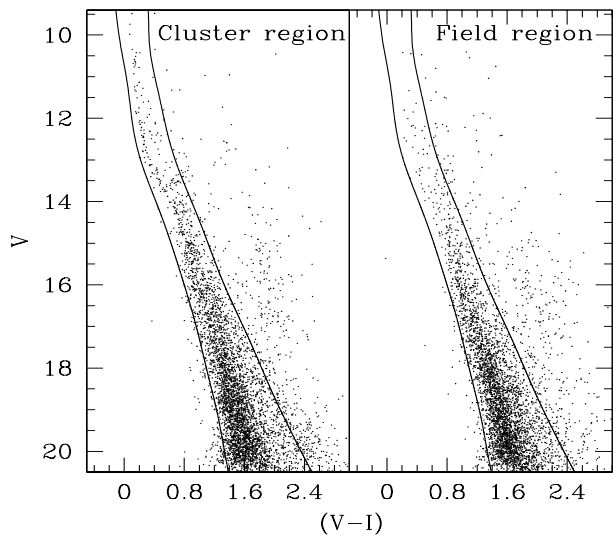}
\caption{$V/(V-I)$ diagrams for stars in the NGC 1960 cluster and field region. The slanted lines
envelop the probable main-sequence stars.}
\end{figure*}

%-------------Fig 3----------------------------------------------------------------
\begin{figure*}
\centering
\includegraphics[height=16cm,width=8cm,angle=-90]{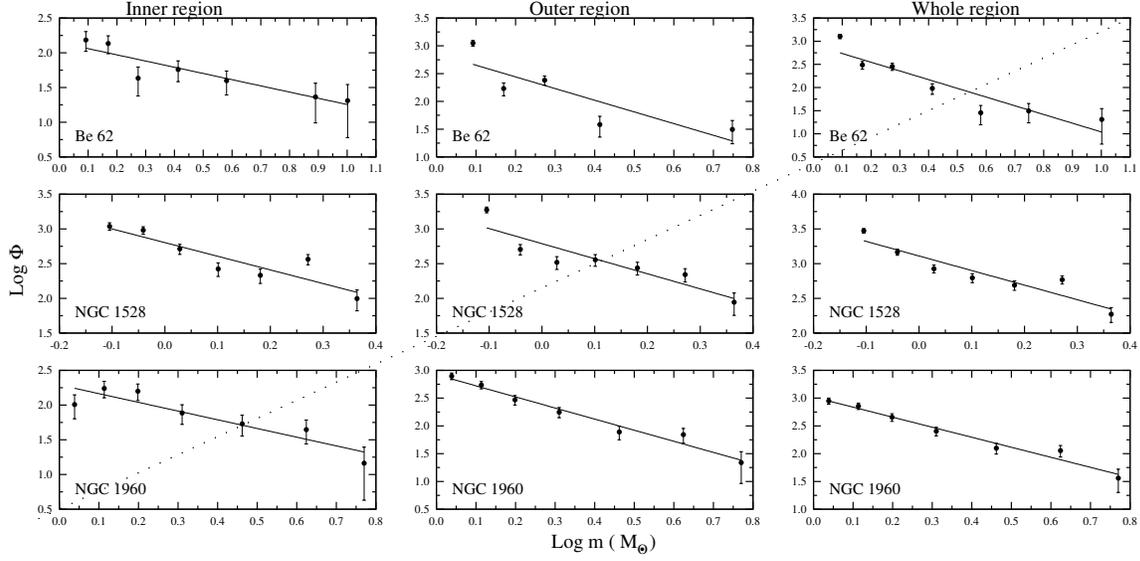}
\caption{ A plot of the mass functions for two sub-regions and the whole-cluster region of the clusters Be 62, 
NGC 1528 and NGC 1960. Log $\phi$ represents log (d$N$/dlog $m$).
The error bars represent $ \pm \sqrt N$
errors. Continuous curves show a least square fit for the given mass range.}
\end{figure*}

\setcounter{figure}{2}

\begin{figure*}
\centering
\includegraphics[height=16cm,width=8cm,angle=-90]{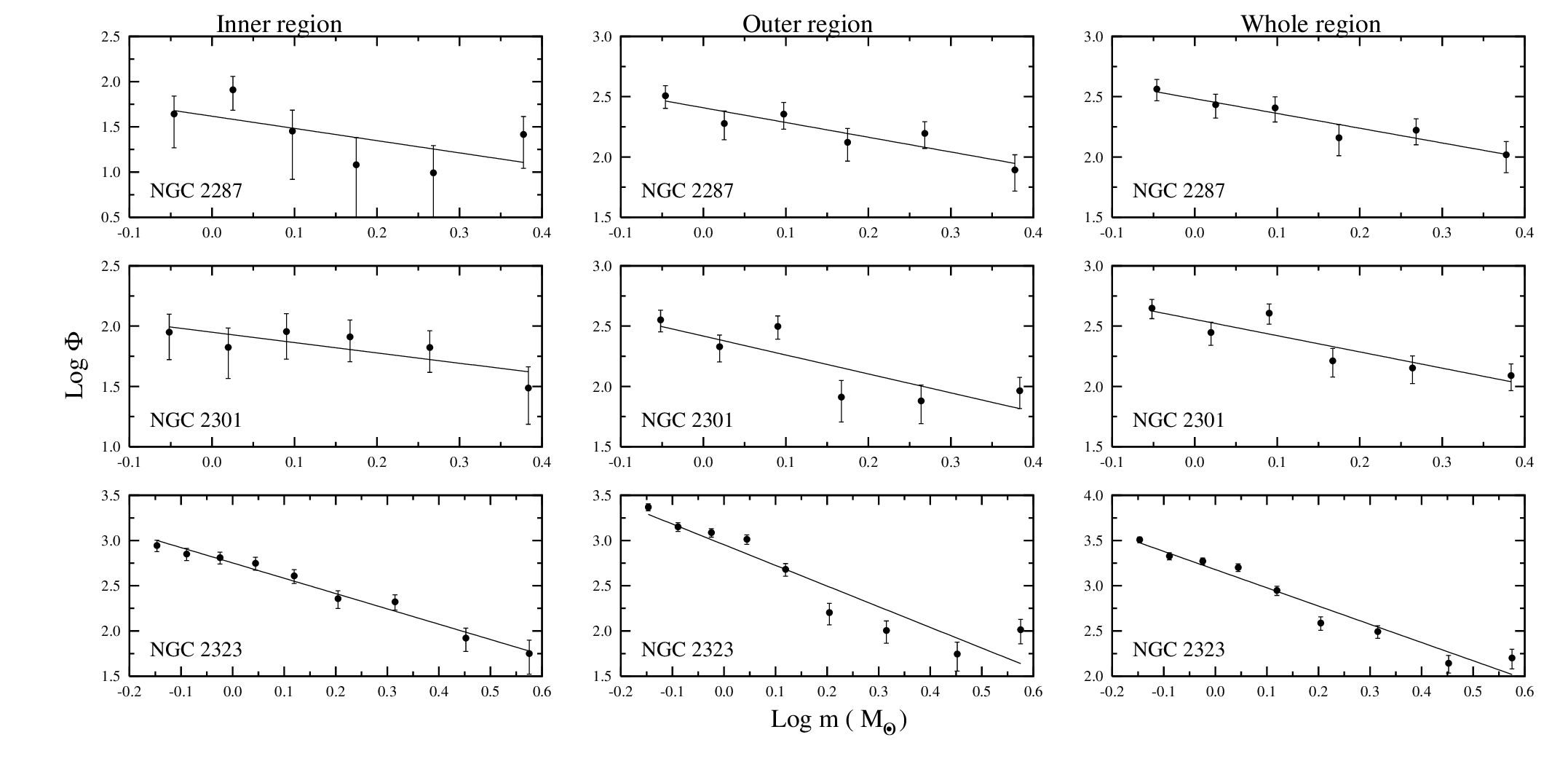}
\caption{Continued - Same as before but for the clusters NGC 2287, NGC 2301 and NGC 2323}
\end{figure*}

\setcounter{figure}{2}

\begin{figure*}
\centering
\includegraphics[height=16cm,width=8cm,angle=-90]{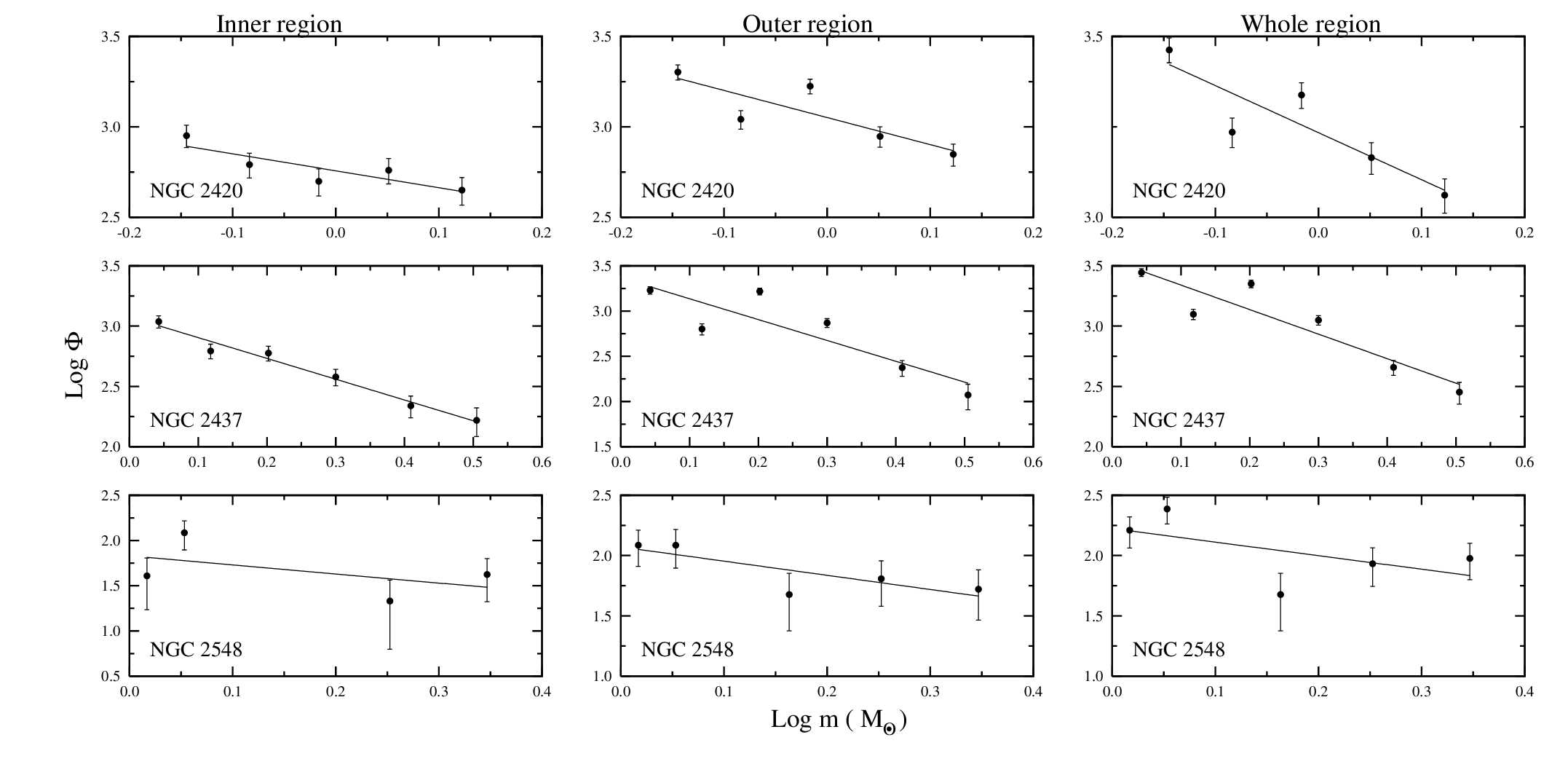}
\caption{Continued - Same as before but for the clusters NGC 2420, NGC 2437 and NGC 2548}
\end{figure*}

%-------------Fig 4----------------------------------------------------------------
\begin{figure*}
\centering
\includegraphics[height=16cm,width=16cm,angle=-90]{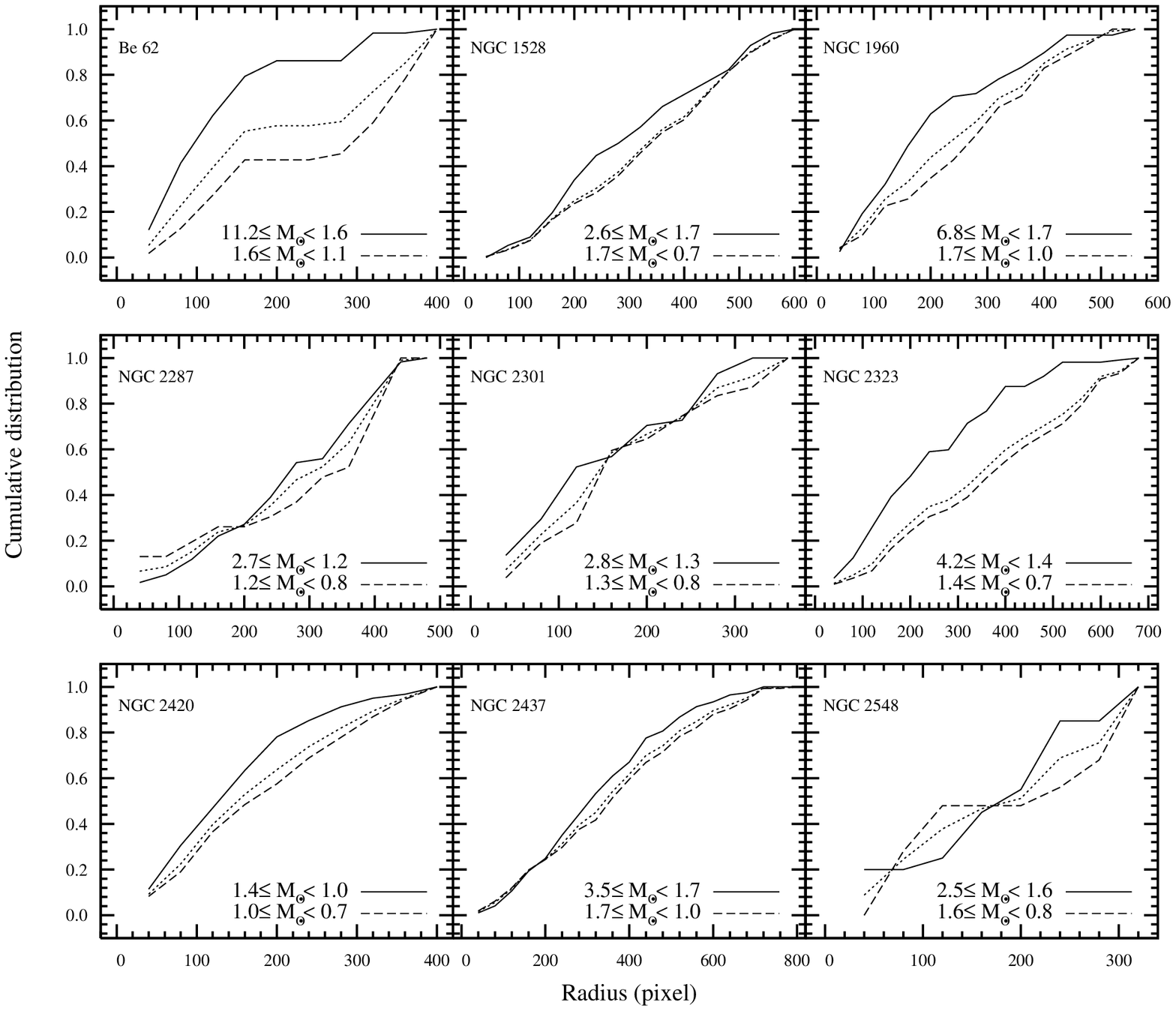}
\caption{The cumulative radial distribution of stars for different mass intervals.
Dotted curves show the cumulative radial distribution of all the cluster  stars.}
\end{figure*}

%-------------Fig 5----------------------------------------------------------------

\begin{figure*}
\centering
\includegraphics[height=14cm,width=8cm,angle=-90]{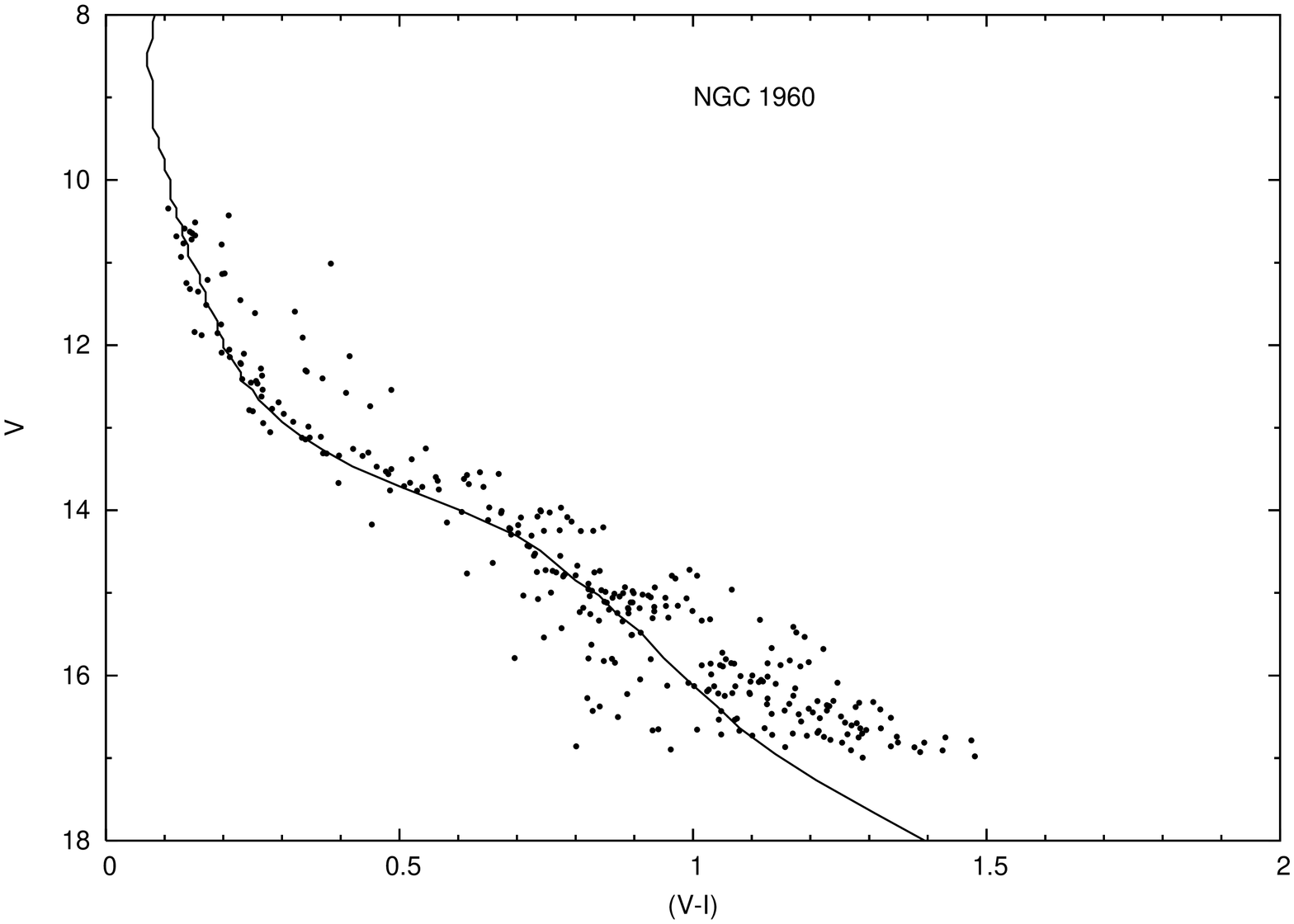}
\caption{Statistically cleaned CMD for the cluster NGC 1960. The solid curve represents the isochrone for 14 Myr for Z=0.02 by Bertelli et al. (1994) adjusted for the distance and reddening of the cluster.}
\end{figure*}

%-------------Fig 6----------------------------------------------------------------
\begin{figure*}
\centering
\includegraphics[height=14cm,width=8cm,angle=-90]{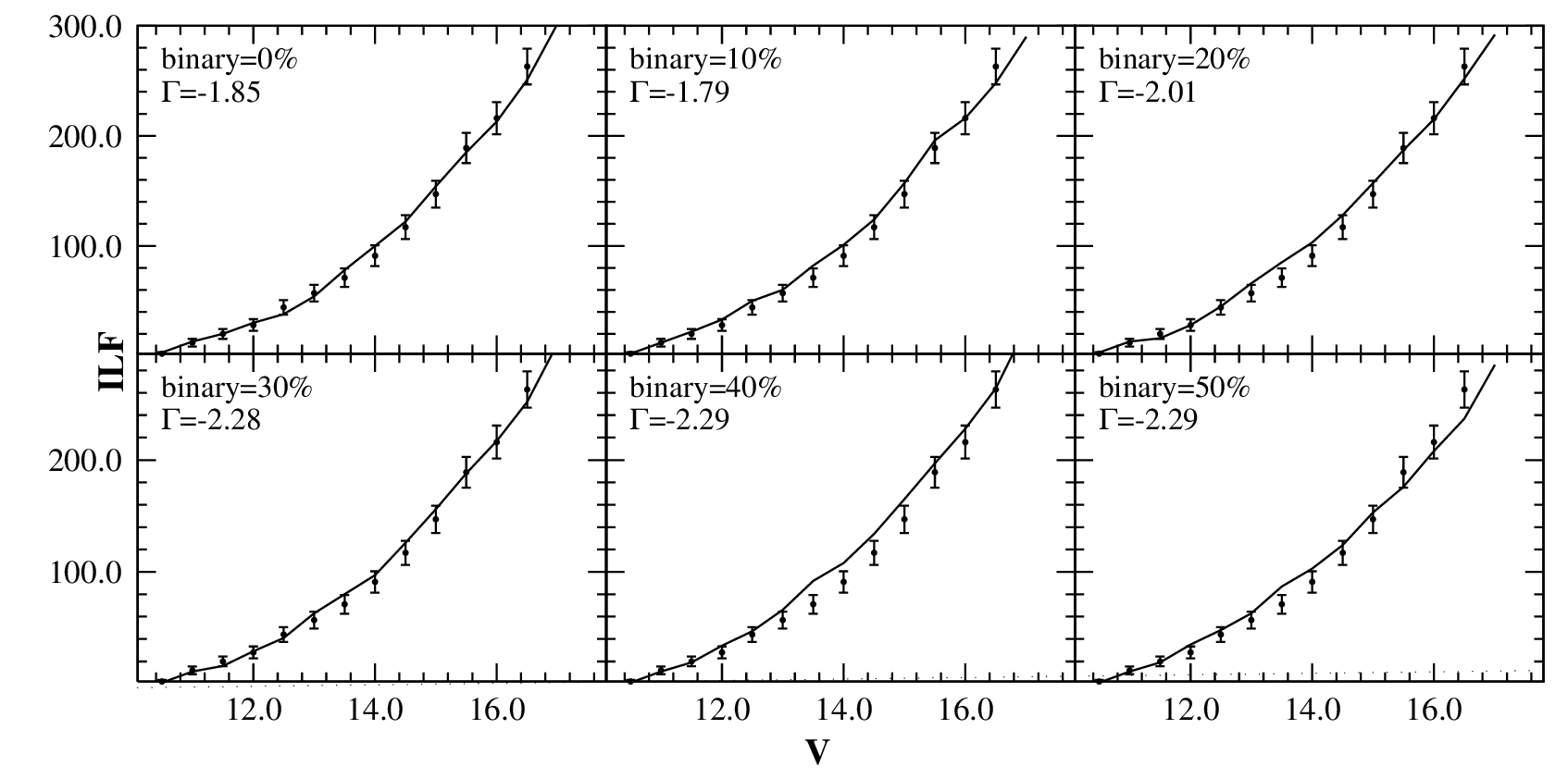}
\caption{Synthetic integrated luminosity functions (ILFs) (continuous curves) of the cluster NGC 1960 are compared with
the observed ILFs (filled circles). The value of binary content assumed and the
values of mass function slopes obtained for the best fit are also given in the figure.
The error bars represent $\pm \sqrt N$ errors in the observations.}
\end{figure*}

%-------------Fig 07----------------------------------------------------------------
\begin{figure*}
\centering
\includegraphics[height=14cm,width=8cm,angle=-90]{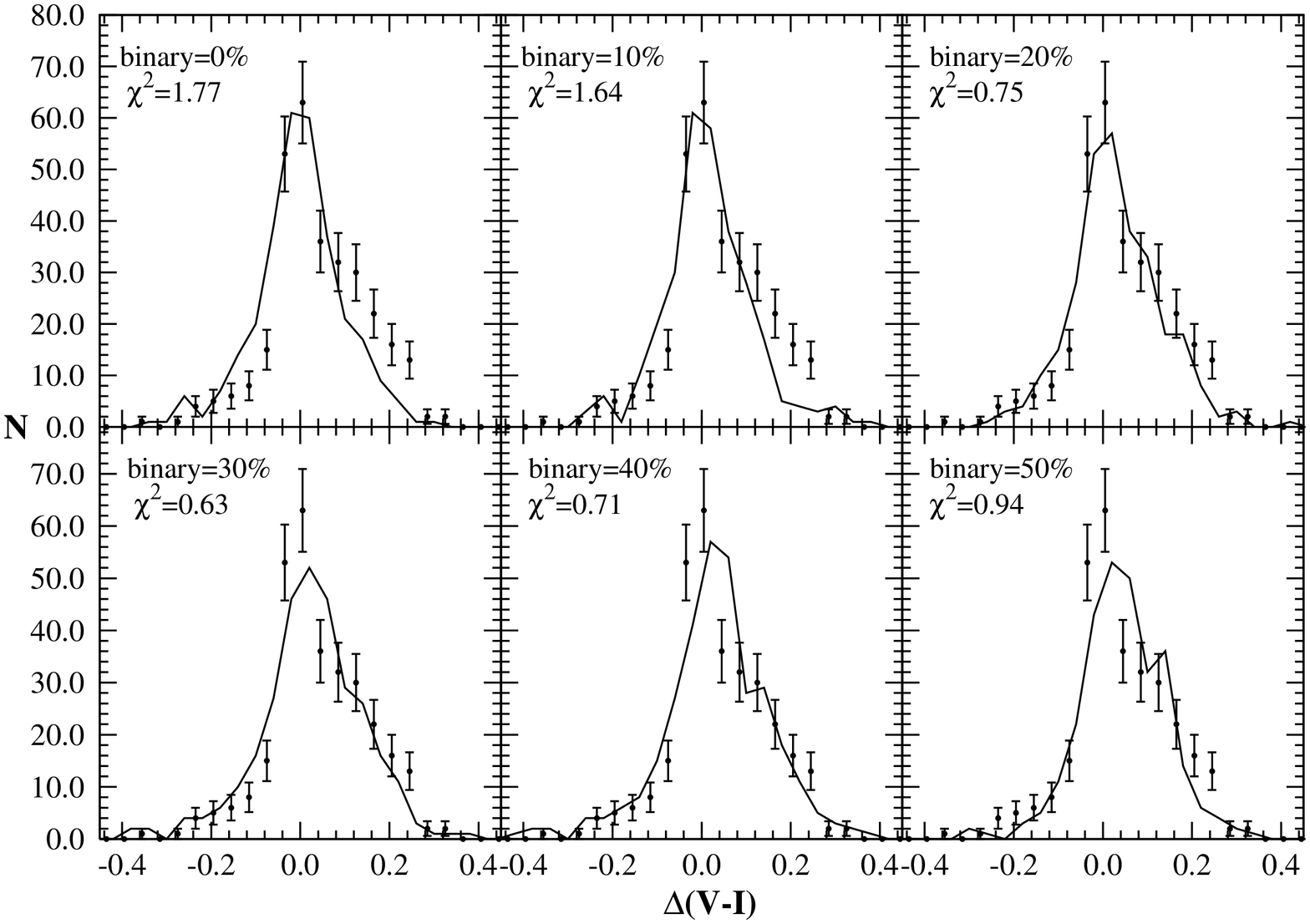}
\caption{Observed (filled points with error bars) and synthetic
$\Delta$(V-I) distributions (continuous curve) for NGC 1960.}
\end{figure*}

%-------------Fig 08----------------------------------------------------------------
\begin{figure*}
\centering
\includegraphics[height=14cm,width=8cm,angle=-90]{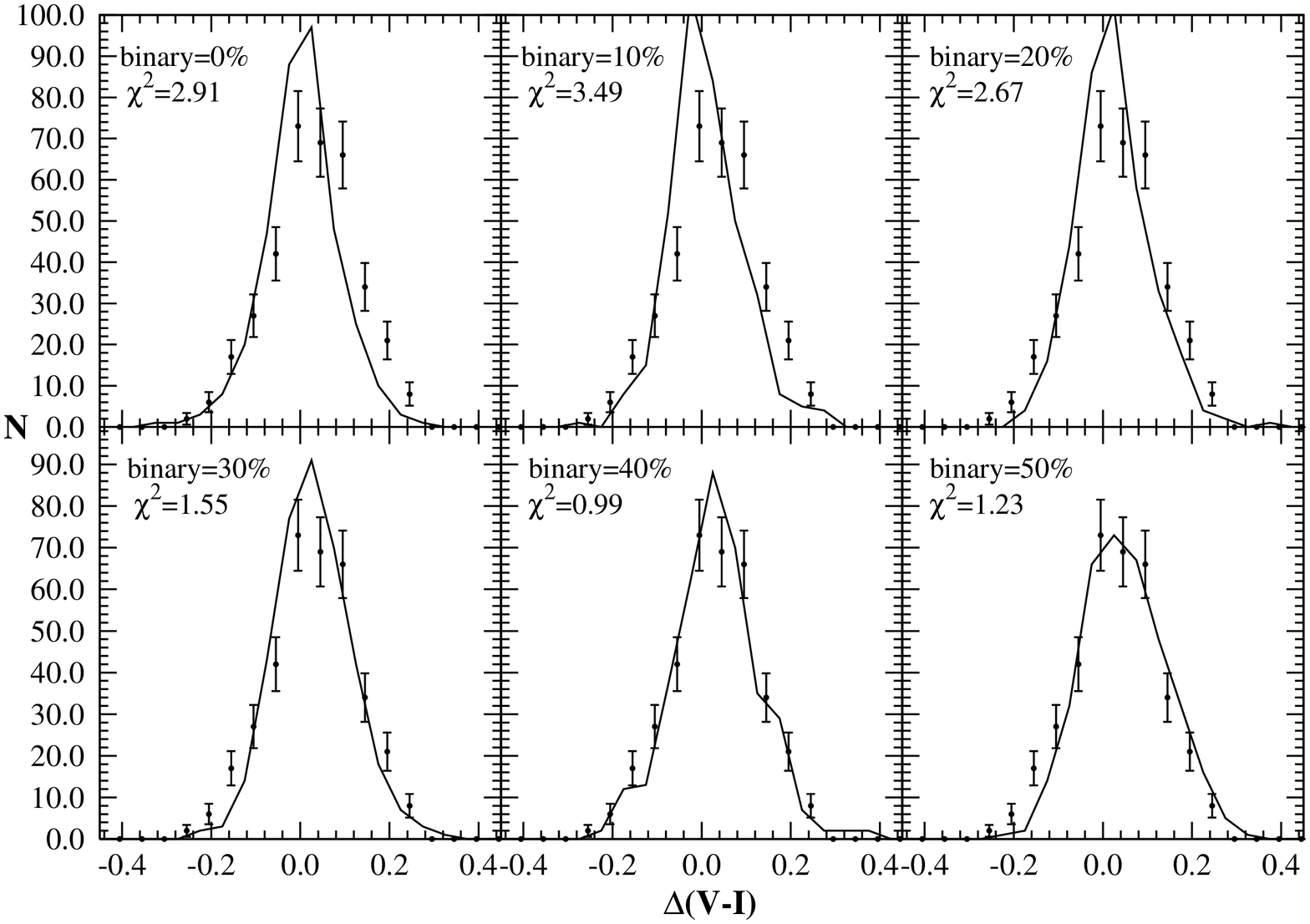}
\caption{Same as Figure 7 but for NGC 1528.}
\end{figure*}

%-------------Fig 09----------------------------------------------------------------
\begin{figure*}
\centering
\includegraphics[height=8cm,width=10cm,angle=-90]{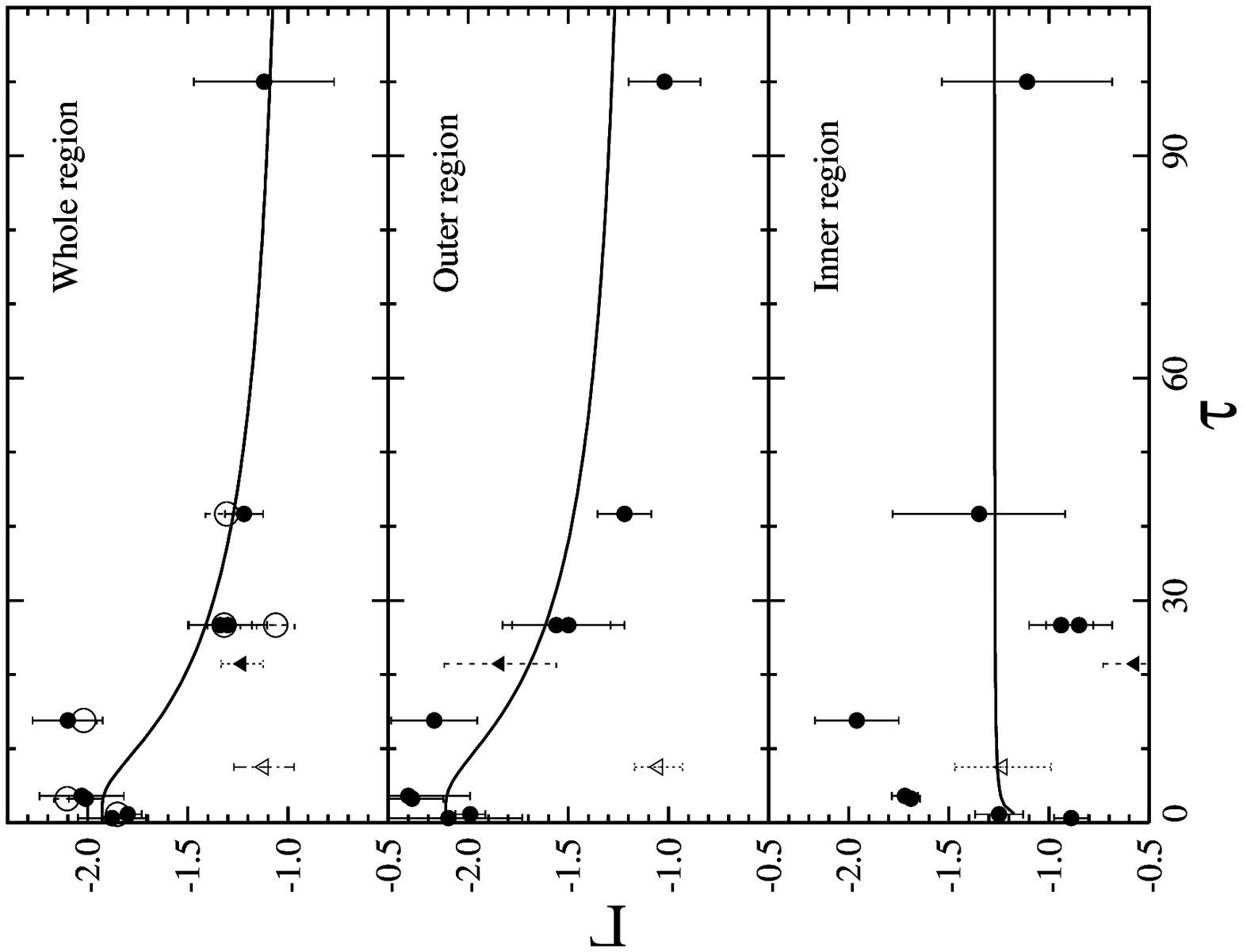}
\caption{Variation of $\Gamma$ as a function of $\tau$ ($age/T_E$) for the whole, outer and inner regions of the clusters.
Filled and open 
triangles represent the slopes of the MFs for the clusters NGC 1907 and NGC 1912 taken from Pandey et al. (2007).
The slope of the MFs for six clusters obtained from synthetic CMDs for 0\% binary are shown by open circles.
Continuous curves show a least square fit to the exponential law (see text). The data point for NGC 1912 (open triangle) is not included in the fit.
}
\end{figure*}

%-------------Fig 10----------------------------------------------------------------
\begin{figure*}
\centering
\includegraphics[height=10cm,width=7cm]{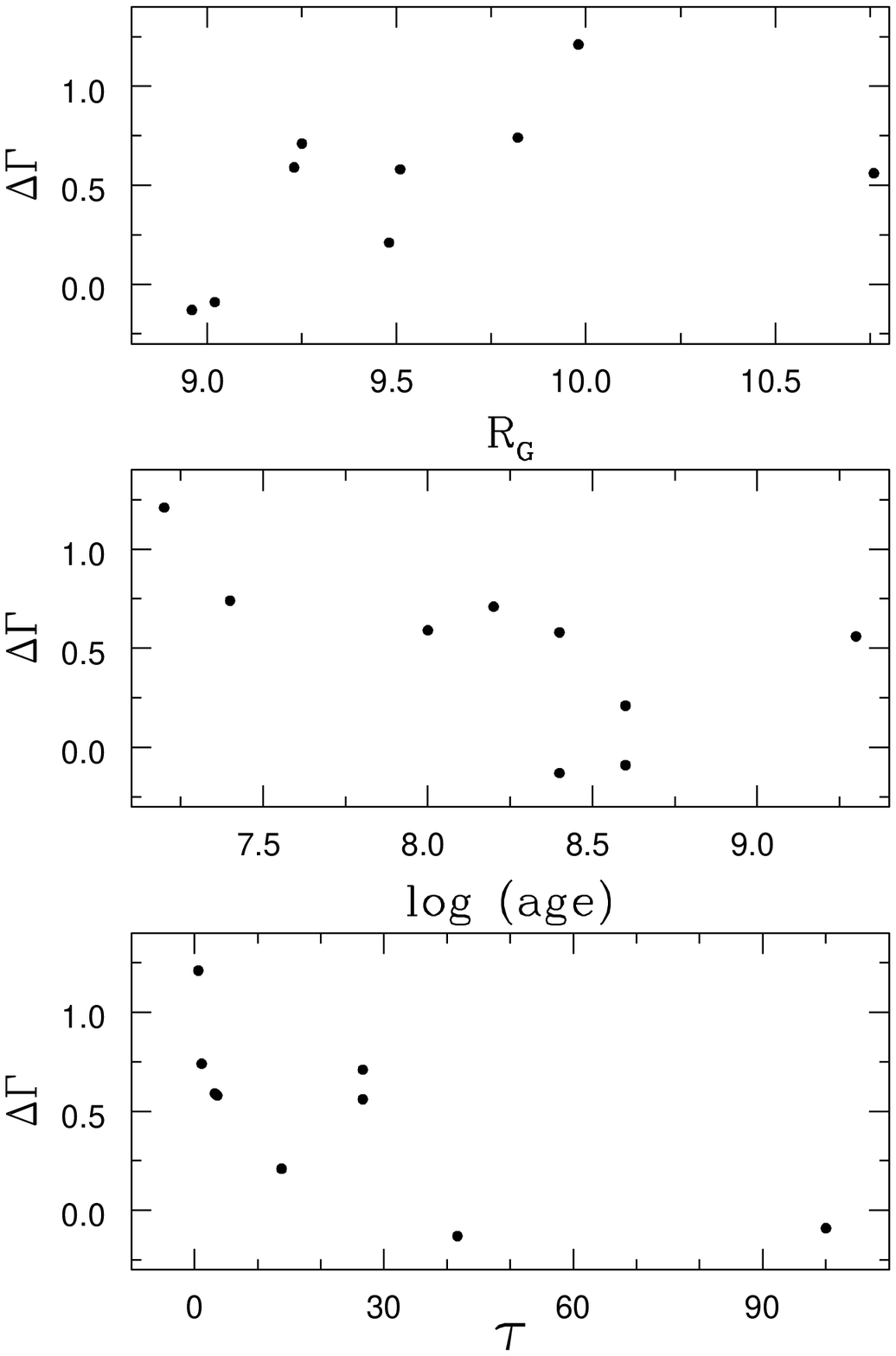}
\caption{ Variation of $\Delta \Gamma$ ($\Gamma_{inner} - \Gamma_{outer}$) as a
function of $\tau$ ($age/T_E$), log (age) and  Galactocentric distance ($R_G$).}
\end{figure*}

%-------------Fig 11----------------------------------------------------------------
\begin{figure*}
\centering
\includegraphics[height=10cm,width=16cm]{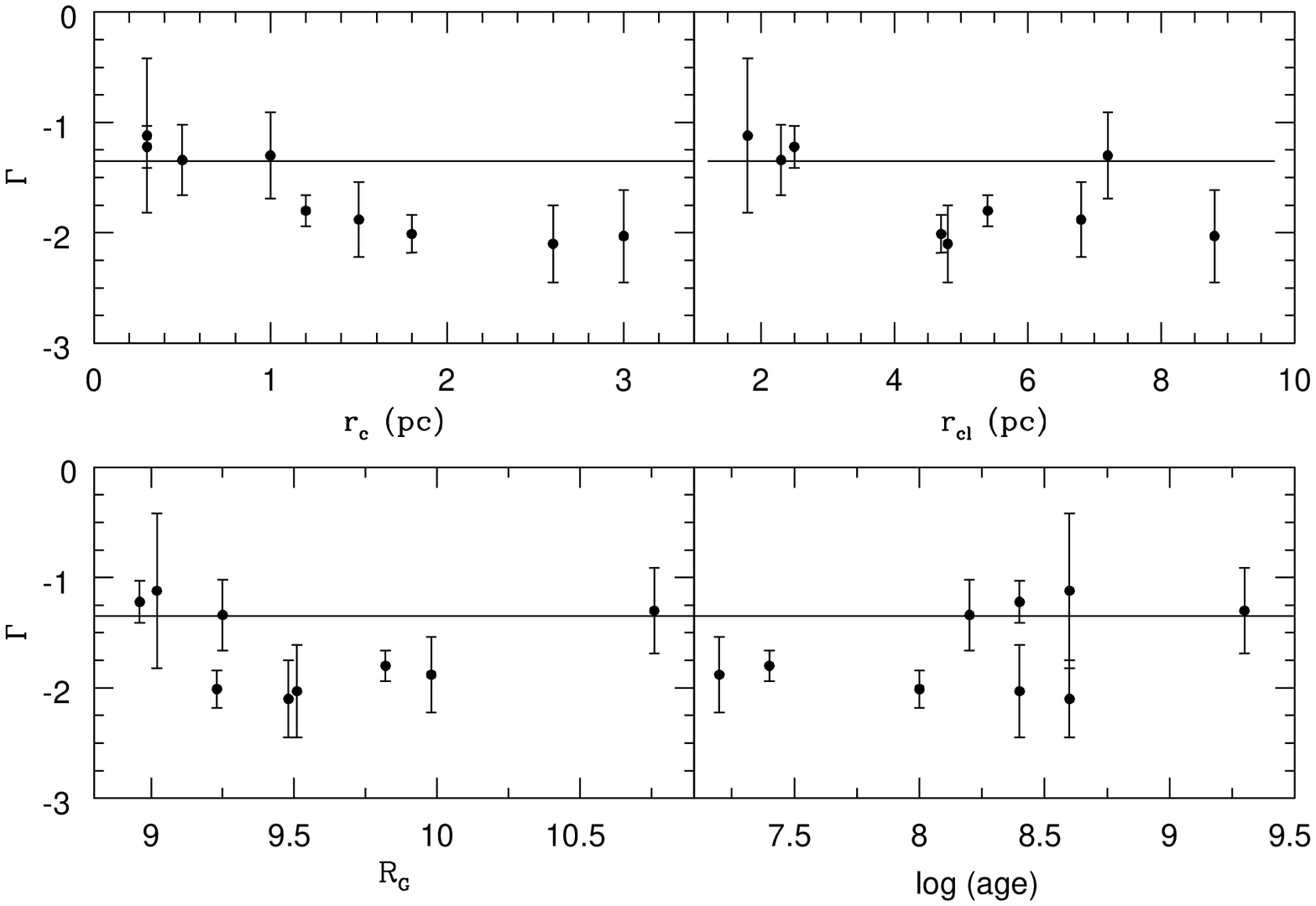}
\caption{Dependence of the MF slopes (whole cluster) on core radius $r_c$, cluster extent $r_{cl}$,
Galactocentric distance $R_G$ and age of the clusters. 
The horizontal lines represents the value of Salpeter slope ($\Gamma = -1.35$).}
\end{figure*}

%\bsp
\label{lastpage}


\begin{thebibliography}{}
\bibitem{}Allen, C. W. 2000, Allen's Astrophysical quantities, London.
\bibitem{}Aparicio, A., Bertelli, G., Chiosi, C., \& Garcia-Pelayo, J. M. 1990, A\&A, 240, 262
\bibitem{}Bertelli, G., Bressan, A., Chiosi, C., Fagotto, F., \& Nasi, E. 1994, A\&AS, 106, 275
\bibitem{}Bonatto, C. \& Bica, E. 2005, A\&A, 437, 483
\bibitem{}Bouvier, J., Rigaut, F., \& Nadeau, D. 1997, A\&A, 323, 139
\bibitem{}Brandl, B., Brandner, W., Eisenhauer, F., Moffat, A. F. J., Palla, F., \& Zinnecker, H. 1999, A\&A, 352, 69	
\bibitem{}Durgapal, A. K., \& Pandey, A. K. 2001, A\&A, 375, 840
\bibitem{}Fischer, P., Pryor, C., Murray, S., Mateo, M., \& Richtler, T. 1998, AJ, 115, 592
\bibitem{}Francic, S. P. 1989, AJ, 98, 3, 888
\bibitem{}Herbst, W., \& Miller, D. P. 1982, AJ, 87, 1478
\bibitem{}Hillenbrand, L. A., 1997, AJ, 113, 1733
\bibitem{}Jeffries, R. D., Thurston, M. R., \& Hambly, N. C. 2001, A\&A, 375, 863
\bibitem{}Kalirai, J. S., Fahlman, G. G., Richer, H. B., \& Ventura, P. 2003, AJ, 126, 1402
\bibitem{}Kholopov, P.N. 1969, SvA-AJ, 12, 625
\bibitem{}King, I.R. 1962, AJ, 67, 471
\bibitem{}Kjeldsen, H., \& Frandsen, S. 1991, A\&AS, 87, 119
\bibitem{}Kroupa, P. 2002, SCIENCE, 295, 82
\bibitem{}Kroupa, P. 2007, arXiv0708.1164
\bibitem{}Kumar, B., Sagar, R., \& Melnick, J. 2008, accepted in MNRAS, arXiv:0801.1068
\bibitem{}Landolt, A.U. 1992, AJ, 104, 340
\bibitem{}Larson, R. B. 1982, MNRAS, 200, 159
\bibitem{}Larson, R. B. 1998, MNRAS, 301, 569
\bibitem{}Lee, S. H., Ann, H. B., \& Kang, Y. W. 2002, ASJ, 273
\bibitem{}Mason B. D., Gies, D. R., Hartkopf, W. I., Bagnudo, W. G., Brummelaar, T. T., \& McAlister, H. A. 1998, 115, 821
\bibitem{}Mathieu, R.D. 1985, IAU Symp., 113, 427
\bibitem{}Mathieu, R.D., \& Latham D.W. 1986, AJ, 92, 1364
\bibitem{}Massey, P., Johnson, K. E., \& DeGioia-Eastwood K. 1995, ApJ, 454, 151
\bibitem{}Maciejewski, G., \& Niedzielski, A. 2007, A\&A, 467, 1065
\bibitem{}McNamara, B.J., \& Sekiguchi, K. 1986, ApJ 310, 613
\bibitem{}Mermilliod, J. C., \& Mayor, M. 1989, A\&A, 219, 125
\bibitem{}Moffat, A. F. J. 1970, PhD thesis, Ruhr Univ. Bochun
\bibitem{}Pandey, A. K., Mahra, H.S., \& Sagar R. 1992, BASI 20, 287
\bibitem{}Pandey, A. K., Nilakshi, Ogura, K., Sagar, R., \& Tarusawa, K. 2001 A\&A, 374, 504
\bibitem{}Pandey, A. K., Upadhyay, K., Ogura, K., Sagar, R., Mohan, V., Mito, H., Bhatt, H.C., \& Bhatt, B.C. 2005, MNRAS, 358, 1290
\bibitem{}Pandey, A. K., Sharma, S., \& Ogura, K. 2006, MNRAS, 273, 255
\bibitem{}Pandey, A. K., Sharma, S., Upadhyay, K., Ogura, K., Sandhu, T. S., Mito, H., \& Sagar, R. 2007, PASJ, 59, 547
\bibitem{}Patience, J., Ghez, A. M., Reid, I. N., Weinberger, A. J., \& Matthews, K. 1998, AJ, 115, 1972
\bibitem{}Phelps, R. L., \& Janes, K. A. 1993, AJ, 106, 1870
\bibitem{}Piskunov, A. E. 1976, Nauch. Inf., 22, 47
\bibitem{}Raboud, D., \& Mermilliod, J. C. 1998, A\&A, 329, 101
\bibitem{}Sagar, R., Piskunov, A. E., Myakutin, V. I., \& Joshi, U. C. 1986, MNRAS, 220, 383
\bibitem{}Sagar, R., Myakutin, V. I., Piskunov, A. E., \& Dluzhnevskaya, O. B. 1988, MNRAS, 234, 831
\bibitem{}Sagar, R., \& Richtler, T. 1991, A\&A, 250, 324
\bibitem{}Salpeter, E. E. 1955, ApJ, 121, 161
\bibitem{}Sandhu, T. S., Pandey, A. K., \& Sagar, R. 2003, A\&A, 408, 515
\bibitem{}Sanner, J., Altmann, M., Brunzendorf, J., \& Geffert, M. 2000, A\&A, 357, 471
\bibitem{}Scalo, J . M. 1986, Fundam. Cosmic Phys., 11, 1
\bibitem{}Scalo, J. M. 1998, in ASP Conf. Ser. 142, The Stellar Initial Mass Function, ed. G. Gilmore \& D. Howell (38th Herstmonceux Conference), 201
\bibitem{}Sharma, S., Pandey, A. K., Ogura, K., Mito, H., Tarusawa, K., \&  Sagar, R. 2006, AJ, 132, 1669
\bibitem{}Spitzer, L., \& Hart, M. H. 1971, ApJ, 164, 399
\bibitem{}Stetson, P. B. 1987, PASP, 99, 191
\bibitem{}Stetson, P. B. 1992, in ASP Conf. Ser. 25, Astronomical Data Analysis Software and Systems I, ed. D. M. Warrall, C. Biemesderfer, \& J. Barnes (San Francisco: ASP), 297


\end{thebibliography}
\end{document}